\documentclass[11pt, letter]{article}

\usepackage[margin=0.7in]{geometry}

\usepackage{microtype}
\usepackage{graphicx}
\usepackage{booktabs} 
\usepackage{hyperref}
\usepackage{todonotes}
\usepackage{amsmath}
\usepackage{tikz}
\usetikzlibrary{plotmarks}
\usepackage{filecontents}
\usepackage{pgfplotstable}
\usepackage{amsfonts}
\usepackage{flushend}

\usepackage{subcaption}
\usepackage{enumitem}
\usepackage{listings}
\usepackage{array}
\usepackage{mathtools}

\usepackage{multirow}
\usepackage{ifthen}
\usepackage{nicefrac}
\usepackage{epigraph}

\newcommand{\namedref}[2]{\hyperref[#2]{#1~\ref*{#2}}}
\newcommand{\sectionref}[1]{\namedref{Section}{#1}}
\newcommand{\figureref}[1]{\namedref{Figure}{#1}}

\newcommand{\Conv}{\mathop{\scalebox{1.5}{\raisebox{-0.2ex}{$\ast$}}}}%
\newcommand{\stimes}{{\times}}

\begin{document}

\title{L3 Fusion: Fast Transformed Convolutions on CPUs}

\newcommand{\footnoteremember}[2]{
\footnote{#2}
\newcounter{#1}
\setcounter{#1}{\value{footnote}}
}
\newcommand{\footnoterecall}[1]{
\footnotemark[\value{#1}]
}

\author{
Rati Gelashvili\footnoteremember{neuralmagic}{NeuralMagic Inc, Somerville, MA}
  \and
Nir Shavit\footnoterecall{neuralmagic}\footnoteremember{MIT}{MIT, Cambridge, MA}
  \and
Aleksandar Zlateski\footnoteremember{FAIR}{Facebook AI Research, New York, NY}\footnoteremember{whileatnm}{Work was done when the author was with NeuralMagic}
}
\date{}
\maketitle

\begin{abstract}

  Fast convolutions via transforms, either Winograd or FFT, had
  emerged as a preferred way of performing the computation of
  convolutional layers, as it greatly reduces the number of required
  operations.  Recent work shows that, for many layer structures, a
  well--designed implementation of fast convolutions can greatly
  utilize modern CPUs, significantly reducing the compute time.
  However, the generous amount of shared L3 cache present on modern
  CPUs is often neglected, and the algorithms are optimized solely for
  the private L2 cache.  In this paper we propose an efficient `L3
  Fusion` algorithm that is specifically designed for CPUs with
  significant amount of shared L3 cache.  Using the hierarchical
  roofline model, we show that in many cases, especially for layers
  with fewer channels, the `L3 fused` approach can greatly outperform
  standard 3 stage one provided by big vendors such as Intel.  We
  validate our theoretical findings, by benchmarking our `L3 fused`
  implementation against publicly available state of the art.

\end{abstract}

\section{Introduction}
  While the Convolutional Neural Networks (\emph{ConvNet}s, or \emph{CNN}s) have been proposed,
  in their current form, at the turn of the century~\cite{lecun99}; it took more than a 
  decade for them to gain a significant traction in the scientific community.  
  The superior performance of AlexNet~\cite{KSH12} in the field of image recognition, together with advances in the computational power of GPUs triggered current popularity of ConvNets.  A whole new field of `Deep Learning`
  has emerged since, studying ConvNets and their applications in various domains.
  
  ConvNets were initially mostly used for research within the academic community and industry research and were running on
  expensive GPU clusters; However, they have since spread to nearly all industries and are running `in production` on
  wide variety on devices -- ranging from traditional GPUs, CPUs all the way to mobile processors and 
  specialized hardware.  As they are computationally very expensive it is very important to have
  optimized algorithms and implementations across large variety of ConvNet architectures and devices in 
  order to achieve satisfactory speeds and save energy.
  
  In this paper we focus on modern CPUs, both server and desktop grade.  Most of the current CPU--based
  algorithms emerged with the introduction of Intel Xeon Phi co--processor, which had many integrated
  x86 cores that provided a competitive computational power to GPUs.  It also provided on--chip high--bandwidth
  memory (MCDRAM), but had limited amount of L2 cache, and completely lacked L3 cache.  In the meantime
  both server and desktop grade CPUs had reached, and even exceeded the computational power of the
  Intel Xeon Phi processors (which quietly got discontinued~\cite{intel-xeon});  however, they lack the high--bandwidth on--chip memory, but instead provide
  a generous amount of L3 cache.

  The most efficient available implementations use `fast convolution algorithms` -- that is, they perform
  convolutions through a basis transform, either FFT or Winograd.  Such algorithms provide a great
  reduction in the number of required floating point operations (FLOPs), but require much more sophisticated 
  design and implementation in order to effectively utilize the underlying hardware.
  
  Nearly all current implementations~\cite{ZJLKD19, falconlib, libxsmmGit, mkl-dnn} of `fast convolutions` do not take
  the L3 cache into account.  While they can still get high utilization in many scenarios on modern CPUs with
  decend amount of L3 cache, there are
  many cases where an L3-cache aware algorithm can significantly boost the performances.  This
  is specifically true for layers of the most popular ConvNets, such as ResNet~\cite{resnet}, or VGG~\cite{vgg}.
  
  In this paper, we propose a L3--cache aware algorithm for the most computationally intensive -- convolutional
  layers.  While the overall idea is quite intuitive -- keeping data reused among cores in L3 cache, the 
  novelty lies within the details of the algorithm.  
  
  We analyze our L3--cache aware algorithm using the roofline model in 
  order to predict in which scenarios it is expected to outperform currently
  available implementations.  We benchmarked our L3 aware implementation against
  the state of the art publicly available implementations, as well as
  our own baseline implementation.   The results conform to our 
  expectations.  As expected, the fused implementation is sometimes slower
  and sometimes faster than the state of the art; while consistently being
  faster by 50\% on average on layers with lower number channels.






\section{Background}
\label{sec:model}

\subsection{Convolutional Layers through Transforms}
Here we'll focus on 2D ConvNets, and follow the notation introduced by~\cite{fbfft}.  The input and the output of a convolutional layer
are $4$-dimensional tensors with shapes of $B \times C \times D \times W$ and $B \times C' \times D' \times W'$ accordingly.  Here $B$ is the batch size, $C$ ($C'$) is the number of input (output) channels, respectively, and $D$ ($D'$) and $W$ ($W'$) are spatial dimensions of each input (output) channel.

We'll further assume isotropic kernels of size $K_D \times K_W = K^2$.  It should be straight forward for the reader to generalize our approach to higher dimensional ConvNets and non--isotropic kernels.  All the kernels thus, have a $4$-dimensional shape of $C' \times C \times K \times K$.

We will assume 32-bit floating point numbers, as it is standard for all
CPU--based implementations of fast convolutional algorithms (using transforms).  FFT operates on complex numbers, which simply requires storing a pair of 32-bit floats.

Convolutional layers optionally have a padding; it typically used to have the input and output of the layer have the same spatial dimensions. Our algorithm can work for any padding, and implementation uses implicit padding which could be more efficient than explicitly padding the data.

  A convolutional layer transforms an input tuple of $C$ images into
  an output tuple of $C'$ images.  A batch of $B$ inputs yielding a
  batch of $B$ outputs is processed at the time via
  { \small
    \begin{equation}\label{eq:forward}
      I'_{b,c'} = \sum_{c=1}^C I_{b,c}\Conv W_{c'c}
    \end{equation}
  }
  
  When using `fast convolutions` (Winograd or FFT), the output images are
  computed via
  \begin{equation}  \small 
    \begin{aligned}
      I'_{b,c'} & = \sum_{c=1}^C \Big[ (W_{c,c'} \stimes_{n=1}^{N}
        \mathbb{G}_n) \odot (I _{b,c} \stimes_{n=1}^{N} \mathbb{B}_n)
        \Big]\stimes_{n=1}^{N} \mathbb{A}_n^T \\ & = \Big[
        \sum_{c=1}^C (W_{c,c'} \stimes_{n=1}^{N} \mathbb{G}_n) \odot
        (I_{b,c} \stimes_{n=1}^{N} \mathbb{B}_n)
        \Big]\stimes_{n=1}^{N} \mathbb{A}_n^T \\
    \end{aligned}
    \label{eqn-layer-comp}
  \end{equation}

  Here, $\odot$ represents element--wise
  multiplication, and the operation $x \times_n^N \mathbb{Y}$ is short for $x
  \times_1 \times_2 \dots \times_n \mathbb{Y}$, where $\times_n$
  represents tensor--matrix mode--n multiplication as defined
  in~\cite{kolda2009tensor, budden}.  $\mathbb{A}_n$, $\mathbb{B}_n$ and $\mathbb{G}_n$ are 
  transform matrices along dimension $n$.  For Fourier transform, complex, matrices, the tensor--matrix
  multiplication can be implemented more efficiently using the FFT algorithm.
  
  For the 2D case, and isomorphic transform sizes, the formula above reduces to
  \begin{equation}  \small 
    \begin{aligned}
      I'_{b,c'} & = \mathbb{A} \Big[
        \sum_{c=1}^C (\mathbb{G} W_{c,c'} \mathbb{G}^T) \odot
        (\mathbb{B} I_{b,c} \mathbb{B}^T)
        \Big] \mathbb{A}^T \\
    \end{aligned}
    \label{eqn-layer-comp-2d}
  \end{equation}
  
  Which was also proposed in~\cite{LS16}.

  Here, the choice of transform matrices assume a particular size of $I$ and $I'$.  
  The Winograd transform matrices produce numerically stable results only for
  relatively small sizes.  The convolution of larger images is performed
  using the overlap--add method (OLA)~\cite{rabiner1975theory}.  With
  OLA, the input images are divided into tiles with sizes of $T$, 
  and an overlap of $K - 1$ along each dimension.
  Considering tiles at the same location from all the input images,
  tiles of size $T' = T - K + 1$ of the output images are computed using the
  formula above.
  
  When FFT transforms are used, the transformed images will be conjugate anti--symmetric.
  This allows for approximate 2x savings in both storage and compute~\cite{ZJLKD19}.
  
  Traditionally, the CPU implementations~\cite{mkl-dnn, libxsmmGit, JZDL18, falconlib} perform
  such computation in a serialized fashion.  First, {\bf all} the inputs and kernels are transformed.  Then,
  the element--wise computation is performed using matrix multiplication; finally, the result is 
  then transformed yielding the result of the convolutional layer.  
  
  
\subsection{Cache hierarchy}
In a typical modern CPU processor there are several layers of cache.  On most Intel processors for example, the third level of cache, known as the L3 cache, is relatively large and shared among all of a on--chip computing cores.  Other level caches, such as L1 and L2, are faster and private to a specific core.

Caching happens at a granularity of cache lines (typically 64 bytes) and if the cached data has a bad alignment in memory, this results in the overhead of unnecessary data being brought to the cache along with the cache lines.  The best (cache-friendly) memory layout is typically storing (and accessing) data consecutively, in cache line size increments, starting from a beginning of a cache line.

Note that the cache coherence protocol is proprietary to the hardware producer, so we cannot guarantee the cache behavior.  However when 
a contiguous chunk of data is frequently accessed (of size smaller than the cache size), we can confidently expect these accesses to be cache hits most of the time for any reasonable cache coherence protocol.

It is worth noting here that while algorithms executed on modern CPUs often implicitly benefit from the existence of the L3 cache. Algorithms that structure the computation to explicitly benefit from the shared cache are not as common. This is because, as opposed to private L2 cache, when multiple processors and the resulting asynchronous nature of computation is involved, it is harder to reliably expect desired data to be in the shared cache. However, in our case, we can structure the transformed convolutions in such a way that all processors repeatedly (in each computation task determined by the algorithm) access the same data, keeping it sufficiently `hot` to avoid eviction from the shared cache;  in addition, a limited amount of additional data is accessed, which further avoids cache pollution.

\subsection{Roofline  Model}

We use Roofline Performance Model~\cite{roofline} to theoretically analyze our algorithms. This model provides a simple, yet a powerful framework for estimating the limit of compute utilization of an algorithm based on the memory movement and the amount of operations performed.  

For caches and memory, the \emph{compute-to-memory ratio} (CMR) is defined as the ratio between the peak theoretical compute performance (typically FLOPS -- Floating operations per second) and the memory/cache bandwidth (bytes/second). Similarly, \emph{Arithmetic intensity} (AI) of an algorithm is the number of compute operations for each byte transferred to or from memory.  When the arithmetic intensity of an algorithm executed on some architecture is not higher than the compute-to-memory ratio of some memory level on that architecture, the execution will necessarily be \emph{memory bound}, i.e. bottlenecked on bringing the data in or out of that memory level at some point in the execution; and the compute resources will have an utilization upper bounded by AI/CMR.

Otherwise, if AI is greater than the architecture's CMR, at a certain level of memory hierarchy, the execution of the algorithm is \emph{compute bound}, i.e. never limited by the memory bandwidth at that memory level.

However, being actually compute bound depends on how the memory accesses are distributed - if the algorithm performs all memory accesses before performing all computation, the average compute utilization of the whole algorithm may be high, despite the first stage of its execution being extremely memory bound. But when an algorithm has a reasonably uniform memory access distribution, then it is likely to utilize a fraction of the CPU’s available FLOPS which is close to its theoretical maximum (i.e. minimum of compute utilization among all memory levels).

\section{Fast Convolutions}
\label{sec:trans}

  Both Winograd and FFT fast convolution algorithms are parameterized by the tile size $T$ they use.  The tile sizes for the
  Winograd are usually small (4--6), as larger tiles yield a numerically unstable results~\cite{LS16, JZDL18}.  All FFT tile sizes
  are generally stable; the sizes are then chosen such that the total number of operations is minimized, and the compute fits
  in caches~\cite{ZJLKD19}.
  
  The output is computed tile-by-tile, where each tile has a shape $T' \times T'$. The output tiles don't overlap, but the pre-image necessary to compute a given output tile consists of $C$ input tiles of shape $T \times T$ aligned across input channels. Moreover, these $C$ input tiles also form the exact pre-image for computing all $C'$ output tiles aligned with the given output tile across output channels. 

  To compute a single tile of size $T' = T - K + 1$ of each of the $C'$ output channels the following steps need to be performed.
  \begin{enumerate}
      \item $C$ input tiles (tiles at the corresponding location in each of $C$ input channels) are transformed.  This yields
      $C$ transformed tiles of size $T^2$, which are then interpreted as $T^2$ vectors of size $C$.
      
      \item Each vector is multiplied by a corresponding matrix of transformed kernels.  There are $T^2$ \emph{right-hand}/\emph{kernel} matrices (one for each location in the input tile, corresponding one-to-one to the vectors) obtained ahead of time by transforming the kernels~\footnote{This is a typical assumption: when performing inference on a trained network, kernels don't change - they can be transformed and stored in a way that is most suitable for the convolution implementation being used. Even for training, there are approaches that store and update transformed kernels, e.g.~\cite{winotrain}.}.

      The output is $T^2$ vectors of size $C'$ that can be interpreted as $C'$ tiles of size $T \times T$.

      \item Each of these $C'$ tiles must finally be \emph{inverse} transformed, resulting in $C'$ output tiles of shape $T' \times T'$. These are the outputs of the convolutional layer at the corresponding locations.     
  \end{enumerate}
  
  The above computation works with all $C$ input tiles and $C'$ output tiles at once. This is the standard way of operation of transformed convolutions. We discuss in~\sectionref{sec:conclusions} whether in some cases, it could be beneficial to break the convolution into multiple convolutions operating on subsets of channels.

  The whole convolutional layer is computed when all output tiles are computed. For this, the above 3-step computation needs to happen $N_{tile}$ times, where $N_{tile} = B \cdot \lceil (D - K + 1)/ T \rceil \cdot \lceil (W - K + 1)/T \rceil$ if we assume no padding.

  As noticed above, the state-of-the-art implementations perform the computation in three fully separate stages.  They perform the first step
  for all the tiles of all batches;  this creates $T^2$ matrices of size $N_{tile} \times C$.  Then the second steps are performed
  by multiplying these $T^2$ matrices with the $T^2$ kernel (right-hand side) matrices, yielding $T^2$ matrices of size $N_{tile} \times C'$.  Finally, all output tiles are computed by transforming the result.

  The second stage operates on $T^2$ matrix pairs of sizes $N_{tile} \times C$ and $C \times C'$ which are typically large and don't fit in cache.  Thus, one matrix multiplication is performed at the time -- total of $T^2$ times.  Note that it is not possible to generate
  only a single matrix at the time, as each transform generates an element from $T^2$ distinct matrices.  Similarly, the output transforms
  can not be performed unless the elements at the same location of all $T^2$ matrices at the same locations were not already computed.
  
  On modern CPUs with large CMRs, the stages 1 and 3 are memory bound as they perform relatively small operations per transferred byte,
  while only the second stage could be compute bound.
  
  This design design is reasonable for architectures with limited L3 cache (or ones with high bandwidth memory), as the $T^2$ kernel matrices can not be stored in cache.
  However, in the case when the $T^2$ kernel (right-hand-side) matrices can fit in shared cache, we can consider performing steps 1--3
  for only a small subset of tiles.  This will allow for larger overall arithmetic intensity, and is the basis of our approach described in the next section.

\section{Algorithm Design}
\label{sec:algo}
Our algorithm is based on a simple, yet crucial observation that the same $T^2$ right-hand (kernel) matrices are used in the second step of the transformed computation, regardless of the output tile. While these matrices may not fit in the smaller L2 private cache of a processor, typically they can comfortably fit (i.e. occupy at most a constant fraction of the total cache size, leaving enough space for other intermediate data used by the processors while executing the algorithm) in the larger shared L3 cache. Fetching memory from the L3 cache is faster than from the main memory, opening an avenue to structure the computation differently from the existing implementations. 


Unlike the state of the art approaches that perform all $N_{tile}$ instances of the $i$-th step in a single $i$-th stage, our algorithm groups $R$ steps together into a task, creating a total of $N_{task} = \lceil N_{tile} / R \rceil$ of tasks.  Each task consists of transforming $R$ groups of $C$ input tiles ($R$ instances of step 1) aligned across the input channel, resulting in $T^2$ left-hand matrices of shape $R \times C$, then multiplying these matrices by the right-hand matrices and inverse transforming the results.

Notice that $\lceil N_{tile} / R \rceil$ tasks can be computed independently of each other, in any order or even in parallel. 
A number of load balancing schemes can be used to schedule/execute 
the tasks on available processors. In any case, every processor fully accesses each right-hand matrix while computing each task, ensuring that these matrices remain extremely ``hot" for caching purposes. Moreover, the right-hand matrices are only read in each task, but never updated (i.e. we don't have to worry about bus traffic for cache coherency). We store the right-hand matrices in a cache-friendly memory layout. Hence, assuming a reasonable cache coherency protocol (and that the size of the matrices is at most a constant fraction of the available L3 cache), we can indeed expect right-hand matrices to be accessed from the shared L3 cache as opposed to from the main memory. 

\subsection{Parameters and Constraints}
\label{sec:constr}
\subsubsection{Number of Channels and the Tile Size}
When choosing the tile size $T$, we have to abide by constraints arising from the nature of the transformed computation itself~\cite{JZDL18, ZJLKD19}. Making tile size larger decreases total area of overlap between input tiles and increases efficiency, but for larger shapes increasing them further brings diminishing returns. FFT and Winograd also have specific constraints~\footnote{FFT operates on complex numbers, while Winograd operates on reals but suffers from numerical precision issues for larger tile sizes}. Based on the existing work, we can say that common $T$ that works well for Winograd is in the $[5, 8]$ range ($7$ or $8$ being better) and $T$ of at least $16$ work well for FFT~\cite{JZDL18, ZJLKD19}

The L3 cache of modern CPUs have a typical size of 1-2MB per core (e.g. on Intel processors 1.375MB for SkylakeX and approximatelly 2.5mb on Desktop Haswell architectures). The right-hand matrices require $CC'T^2$ numbers to be stored;  this adds to $4CC'T^2$ for both Winograd and FFT~\footnote{While the FFT needs to store complex numbers, that require a pair of floats per number, it only has to store half of the matrices, due to conjugate anti-symmetric nature of the FFT tranforms.} FFT with $T=16$ would require $1$MB total for 32 input and output channels and $4$MB total for 64 input and output channels. Winograd with $T=8$ would require $4$MB even with 128 input and output channels. For more cores (e.g. 18), we could go to executing our algorithm on up to $128$ channels for FFT and $256$ channels for Winograd.

Hence, we can set $C, C'$ and $T$ such that right-hand matrices fit in the L3 cache and tiles make sense from the transformed computation prospective. In the following, we focus our attention on this meaningful setting of parameters and consider the constraints and implications for the remaining key parameter $R$. 

The main, non-trivial question for the remaining of the paper is whether the parameter $R$ can be chosen, and the actual details of the algorithm designed in such a way that it is more efficient than the 3-stage approach. We answer this question in the affirmative from both theoretical and practical prospective.

\subsubsection{Setting of $R$}
The implementation of the forward transform can utilize streaming instructions for reading the input tiles, and similarly, the inverse transform can utilize streaming writes for output tiles. Right-hand matrices can be streamed from the L3 cache for multiplication. However, we need to ensure that during the matrix multiplications, left-hand matrices and the results of the multiplication do not fall out of the L2 cache. More precisely:

\textbf{Lower Bound:} $R$ has to be large enough that the matrix multiplications have a large arithmetic intensity to give a good compute utilization, given that the right-hand matrices are read from the L3 cache, while input tiles are read from and output tiles are written to the main memory.

\textbf{Upper Bound:} $R$ has to be such that the left-hand matrices together with the resulting matrices fit into L2 cache, so that (a) left-hand matrices are in the L2 cache while the results of the multiplication are computed and (b) the results are read from the L2 cache by the subsequent inverse transform.

Notice that violating the upper bound means that we may have to access main memory (or sometimes, L3 cache) instead of L2 cache for crucial intermediate data, which is likely to greatly and adversely affect the performance. Therefore, we treat the upper bound as a hard constraint. On the other hand, lower bound is a softer constraint. There is no single threshold that has to be satisfied - instead there is a range where increasing $R$ improves the compute utilization of the matrix multiplications.

Generally we want $R$ as large as possible, but not higher than the upper bound. We perform the roofline analysis in~\sectionref{sec:analysis}. But first, we introduce a technique that is critical for the efficiency of our algorithm. Without this technique, the upper bound constraint would force us to pick smaller $R$, which might limit us in achieving a good arithmetic intensity. With this technique, we can fit more data in L2 cache, relaxing upper bound constraint almost by a factor of two. This lets us pick a larger $R$ and directly translates into better compute utilization.

\subsection{Shared Buffer}
Recall that the left-hand matrices have shape $R \times C$ and the result matrices of the multiplication have shape $R \times C'$. There are $T^2$ pairs of these matrices. As the upper bound constraint in the previous section dictates, we would like to keep all these matrices at (no lower than) the L2 cache of the processor performing the task. In particular, we would like the left-hand matrices to live in the L2 cache from the moment when they are generated by the forward transform to the moment when they are accessed for the matrix multiplication purposes. Similarly, we would like the matrix multiplication results to live in the L2 cache by the time of the inverse transform accesses.

The processor is performing $T^2$ matrix-matrix multiplications whereby the right-hand matrices are streamed from the L3 cache. So, a standard way for us to satisfy the constraint is to pick $R$ such that the total amount of memory required for both left-hand and result matrices, $T^2 \cdot (4RC + 4R C') = 
4 R T^2 \cdot (C + C')$ bytes, fits in a constant fraction (typically 50\%) of the processors L2 cache. The shared buffer allows us to use significantly less space, $T^2 S_{max} + S_{min}$ to be precise, where $S_{max} = \max(4RC, 4RC')$ and $S_{min} = \min(4RC, 4RC')$. $S_{max}$ and $S_{min}$ represent the the maximum and minimum, respectively, between the memory requirement of a left-hand matrix and memory requirement of a result matrix.

\begin{figure*}
\begin{subfigure}[b]{0.45\textwidth}
        \subcaption[short for lof]{$S_{min} = S_{max} = 32$ bytes}
\includegraphics[scale=0.38]{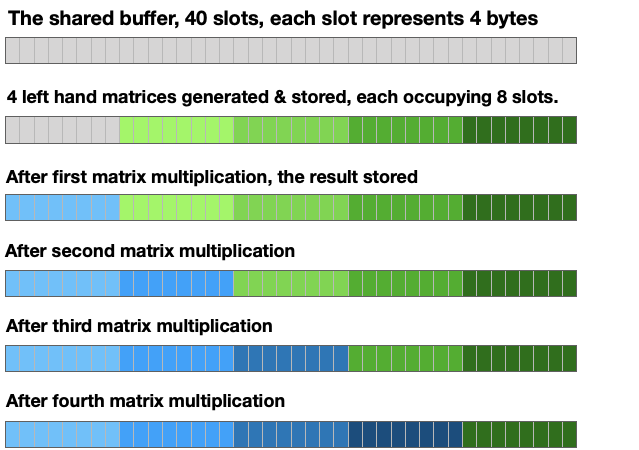}
    \end{subfigure}%
    \hspace{20pt}
\begin{subfigure}[b]{0.45\textwidth}
        \subcaption[short for lof]{$S_{min} = 24$ bytes and $S_{max} = 40$ bytes}
\includegraphics[scale=0.38]{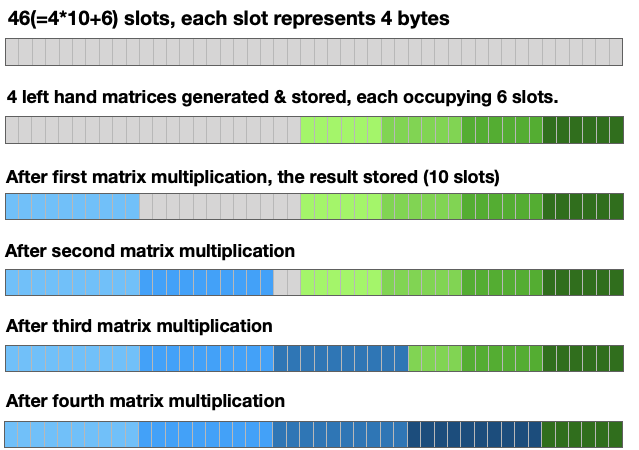}
    \end{subfigure}%

\caption{Simple examples with 4 matrix multiplications. (a): each left-hand matrix and the result matrix has size 32 bytes, occupying 8 slots (e.g. 4-byte floats). The memory storing left-hand matrices are colored in shades of green, and the memory storing the results of multiplications are colored in shades of blue. The buffer occupies 40 slots (160 bytes) providing an 37.5\% improvement versus storing these matrices separately, which would require $32 \cdot 2 = 64$ slots (256 bytes). (b) in this example, left-hand matrices occupy 24 bytes (6 slots) while results of multiplication occupy 40 bytes (10 slots) each. Here shared buffer provides 28.125\% savings (46 slots vs 64 slots).}
\label{fig:simplebuf}
\end{figure*}

\figureref{fig:simplebuf} illustrates the savings of memory in the L2 cache using the shared buffer. The left-hand matrices are stored consecutively aligned with the end of the buffer, leaving some extra unused space in the beginning of the buffer. In example (a) $S_{max} = S_{min}$ (i.e. $C = C'$), so the result of the first matrix multiplication is written precisely to the extra space at the beginning of the shared buffer, while the result of the first matrix multiplication is stored precisely in place of the first left-hand matrix, which is no longer needed. The key aspect is that once $i$-th multiplication happens, the $i$-th left-hand matrix is no longer needed, and its space can be reused for the subsequent result matrices.

The scheme is general and works for arbitrary sizes. In fact, when the size of a result matrix is smaller or equal than the size of a left-hand matrix, at the end of the all $T^2$ multiplications the shared buffer contains all result matrices, followed by the last left-hand matrix. \figureref{fig:simplebuf}-(b) provides a more general example with this property. 

In general, the results of the $i$-th multiplication may overwrite contents of up-to $(i-1)$-st left-hand matrices, but never the $i$-th left-hand matrix (or later ones), as matrix multiplication can not be efficiently performed in--place\footnote{For efficient implementation, an input of matrix multiplication may be read in any order or more than once, and cannot be overwritten.}. However, overwriting left-hand matrices used by completed (earlier) multiplications is always safe.
\section{Theoretical Analysis}
\label{sec:analysis}
In~\sectionref{sec:constr} we established that for reasonable tile sizes for the transforms, all right-hand matrices can comfortably fit in the modern CPUs' L3 cache of modern processors if we consider up to 128 or 256 (64 or 128), input and output channels for Winograd (or FFT), respectively.

However, we still need to ensure that we can pick an $R$ that (a) gives sufficient compute utilization for our algorithm given all the required L3 and main memory accesses, and (b) simultaneously, allows shared buffer to fit in a constant fraction (around 50\%)~\footnote{Typical design choice.} of a processor's L2 cache.

\subsection{Arithmetic Intensity vs CMR}
For the purposes of this section, we assume that the intermediate data while performing tasks do not spill beyond the L2 cache. This is a hard requirement which we consider below. 

The arithmetic complexity of each task is at least $\alpha 2RCC'T^2$, which is the number of FLOPs required to perform the matrix-matrix multiplications after the forward transform.  Here $\alpha$ is 1 for Winograd and $2$ for FFT~\cite{ZJLKD19}.

The amount of memory that is read from the L3 cache is $4C C'T^2$. This gives a CMR of $R/2$ for the L3 cache.  While there's no available specification of
the L3 bandwidth on Intel processors, we know that they operate on cache--ring frequency, and are generally constant regardless of the number of cores.  The L3's CMR should, therefore be obtained empirically by measuring the throughput of the L3 cache, and dividing the theoretical peak FLOPS of the chip by the obtained number.  For the two machines we used in the experimental section, we get that the L3 CMR of the SkylakeX processor was around 10, and for the Kaby Lake mobile processor (i7 Macbook Pro) it was around 4~\footnote{The actual requirements were a bit lower, but we would like to take a conservative estimate.}.

Hence, on the SkylakeX processor, we need $R \geq 20$, and for the mobile i7 CPU we need $R \geq 8$ in order to aim for full compute utilization at the L3 level.

For the main memory, the CMR can be easily computed as the ratio of the processors peak FLOPS and the memory bandwidth (number of channels times the frequency times 8 bytes per transfer).  Which was 35 for the SkylakeX and 13 for the i7.  The size of the input is $4RT^2 C$, the size of the output is $4RT^2 C'$, and the arithmetic complexity is at least $2R C C' T^2$. Hence, when the input tiles are read from the main memory, and the output tiles are written to the main memory, the compute utilization would be $\frac{C C'}{2 \cdot (C + C')} \geq \min(C, C') / 4$. Hence, when we have at least $60$ input and output channels for SkylakeX architectures (and at least $24$ for i9), it is also possible to achieve full utilization at the main memory level.
At least $32$ or $64$ channels is in line with our previous constraints (i.e. we can afford more channels and still fit the right-hand matrices it in L3 cache).

\subsection{Fitting into L2}
Recall that the shared buffer reduces our memory requirement to $4RT^2\max(C, C') + 4R\min(C, C')$ bytes. This amount is upper bounded by $4R\max(C, C') \cdot (T^2 + 1)$ which is quantitatively nicer to work with (and may also allow for a more natural implementations).

The size of L2 cache size is typically 256kb bytes for AVX2 (our i7) architectures, and 1mb for AVX512 (SkylakeX) architectures. We would like the shared buffer to fit in half of the available L2 cache, and using our upper bound we get that, for i7, $$R\max(C, C') \cdot (T^2 + 1) \leq 256 \cdot 1024 / (4 \cdot 2) = 32 \text{kb}$$ 
needs to hold. Analogous derivation for SkylakeX gives the following requirement
$$R\max(C, C') \cdot (T^2 + 1) \leq 128 \text{kb}.$$

\section{Experiments}
\label{sec:exp}

We executed the main experiments on an $18$ core Intel 7980xe
with $2.6$ghz 4 memory channels each $21.3$gb/sec, $20$mb shared L3 cache and $1$mb L2 per-core cache. Despite being a desktop grade machine, it provides native AVX512 instruction set support.

We compare our implementation of L3-fused Winograd convolution to the state-of-the-art implementations of ZNN~\cite{JZDL18} and DNNL~\cite{mkl-dnn} (formerly known as MKL-DNN), along with our own non-fused 3-stage implementation. We run experiments for typical convolutional layers in the VGG and ResNet networks, $4$ layer for each, ranging from $64$ to $512$ channels. The difference is that VGG's layers have $4$ times larger spatial dimensions. In particular, VGG layers that we experiment on are 
\begin{itemize}
    \item $64$ channels and $D=W=224$,
    \item $128$ channels and $D=W=112$,
    \item $256$ channels and $D=W=56$, and
    \item $512$ channels and $D=W=28$.
\end{itemize}
\noindent ResNet layers that we experiment on are:
\begin{itemize}
    \item $64$ channels and $D=W=56$,
    \item $128$ channels and $D=W=28$,
    \item $256$ channels and $D=W=14$, and
    \item $512$ channels and $D=W=7$.
\end{itemize}

All these layers have kernel size $3 \times 3$ and low and high padding of $1$. We choose batch $64$ which is typical for benchmarks. Together, these layers cover typical combinations of possibilities for a range of channels and spatial dimensions.

We let ZNN optimize its parameter space for each layer (in particular, we let it choose different tile size and other internal parameters, such as row blocking, specifically, for each one of the $8$ layer types tested). While DNNL does not expose an interface for parameter search, we also benchmark layer types independently, allowing the framework to make the best decisions for each one. We also pick the best times for each layer obtained by our non-fused implementation among different configurations. Our baseline non-fused implementation is not particularly efficient (nor is that the point of the paper), but it does get competitive results to ZNN and DNNL for most of the tested layers.

On the other hand, we fix a reasonable configuration for the L3-fused algorithm, with $R = 24$ and $T = 7$ (i.e. $7 \times 7$ tile size), in line with our theoretical derivations, and run the algorithm with this fixed configuration for all $8$ tested layer types. This way, we demonstrate the robustness of our algorithm, and it's superior performance even without fine tuning parameters for each particular layer (which can only improve the results further). 

\begin{figure*}
\begin{subfigure}[b]{0.5\textwidth}
        \label{fig:subfigresnet}

\begin{tikzpicture}
\begin{axis}[
    title={ResNet Layers},
    xmode=log,
    log ticks with fixed point,
    scaled x ticks=real:2,
    xlabel={Number of Channels},
    ylabel={Execution Time (ms)},
    xmin=0, xmax=600,
    ymin=0, ymax=17,
    xtick={64,128,256,512},
    ytick={0,4,8,12,16},
    legend pos=north west,
    ymajorgrids=true,
    grid style=dashed,
    legend style={at={(0.5,-0.2)},anchor=north}
]

\addplot[
    color=blue,
    mark=square,
    ]
    coordinates {
    (64, 11.6196)(128, 5.8354)(256, 4.1756)(512, 4.6406)
    };
    ]
 
 \addplot[
    color=red,
    mark=triangle,
    ]
    coordinates {
    (64, 5.3828)(128, 4.2234)(256, 7.6254)(512, 6.0619)
    };
    ]

 \addplot[
    color=black,
    mark=*,
    dotted,
    ]
    coordinates {
    (64, 13.01)(128, 6.436)(256, 4.356)(512, 5.976)
    };
    ]

 \addplot[
    color=green,
    mark=*,
    thick,
    ]
    coordinates {
    (64, 3.155)(128, 2.821)(256, 4.949)(512, 9.764)
    };
    ]

\legend{ZNN, DNNL, our non-fused, L3 fused}
\end{axis}
\end{tikzpicture}
\end{subfigure}%
\hspace{10pt}
\begin{subfigure}[b]{0.4\textwidth}
\begin{tikzpicture}
\begin{axis}[
    title={VGG Layers},
    xmode=log,
    log ticks with fixed point,
    scaled x ticks=real:2,
    xlabel={Number of Channels},
    ylabel={Execution Time (ms)},
    xmin=0, xmax=600,
    ymin=0, ymax=250,
    xtick={64,128,256,512},
    ytick={0,50,100,150,200,250},
    legend pos=north west,
    ymajorgrids=true,
    grid style=dashed,
]

\addplot[
    color=blue,
    mark=square,
    ]
    coordinates {
    (64, 170.876)(128, 88.3928)(256, 61.989)(512, 45.1622)
    };
    ]
 
 \addplot[
    color=red,
    mark=triangle,
    ]
    coordinates {
    (64, 73.0406)(128, 57.9966)(256, 92.539)(512, 59.6532)
    };
    ]

 \addplot[
    color=black,
    mark=*,
    dotted,
    ]
    coordinates {
    (64, 203.17)(128, 116.7)(256, 66.27)(512, 51.39)
    };
    ]

 \addplot[
    color=green,
    mark=*,
    thick,
    ]
    coordinates {
    (64, 46.27)(128, 41.62)(256, 48.97)(512, 93.06)
    };
    ]


\end{axis}
\end{tikzpicture}
        \label{fig:subfigvgg}
    \end{subfigure}%
\caption{Benchmark results on 18-core Intel 7980xe}.
\label{fig:18core}
\end{figure*}
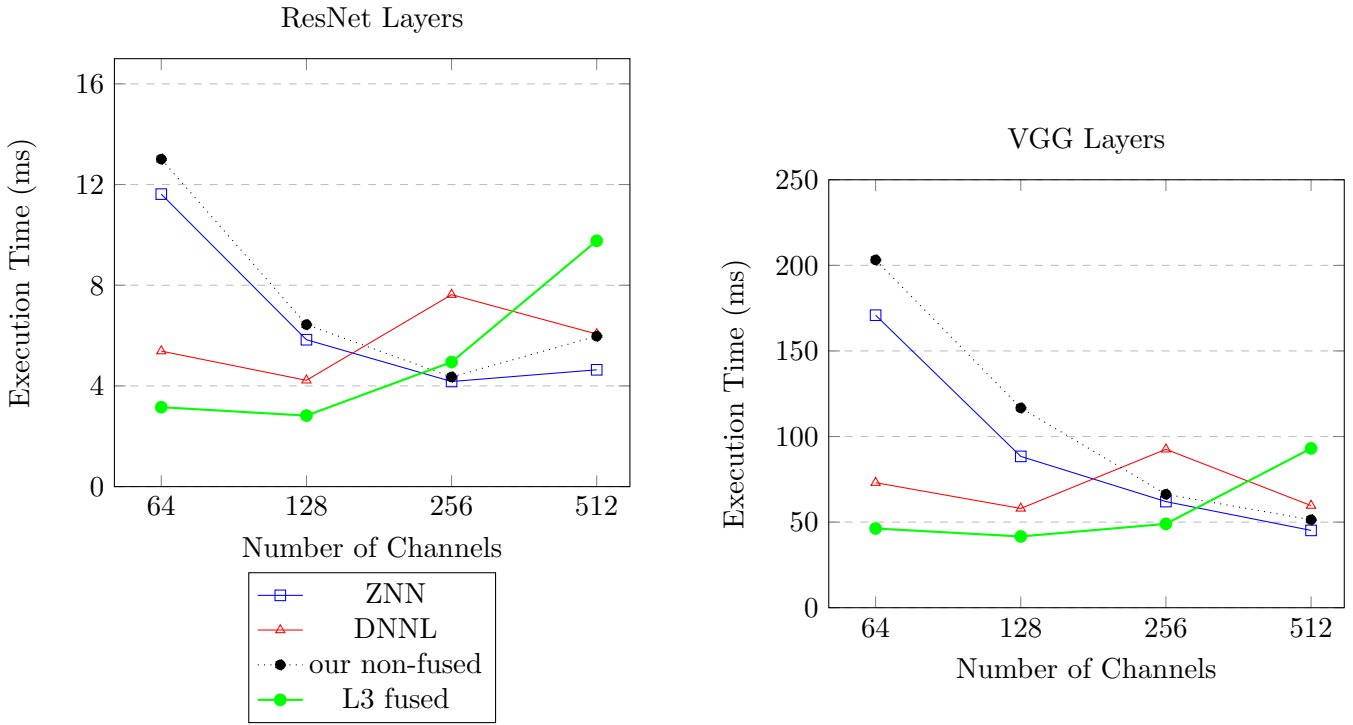

The results are shown on~\figureref{fig:18core} and are consistent with our theoretical predictions. L3-fused algorithm reliably and significantly outperforms the best of all 3 other implementations on all layers with 64 and 128 channels. On 64 channel layer of ResNet (VGG), L3 fusion takes $3.16$ ($46.27$) ms as opposed to the second best time of $5.38$ ($73.04$) ms of DNNL. The improvements are similar for $128$ channels, while on $256$ channels L3-Fusion still achieves close to the best performance in both cases (actually the best performance on the VGG layer).

\subsection{i7 experiment}
We also perform experiments on 4-core MacBook Pro with 3.1 GHz Intel Core i7, 1.6ghz 2 memory channels each 12.8 gb/s, 
$8$ mb shared L3 cache and $256$kb L2 per-core cache. AVX2 instruction set is natively supported, but avx512 instruction set is not supported.

We compare L3 fused algorithm with our baseline (DNNL and ZNN only support CPUs with avx512 instruction set), for tile size $7$.
Based on our calculations, we set parameter $R=8$ for the L3-fused algorithm. Since we expected the performance of the L3-fused algorithm to be the best on i7--like architectures for smaller number of channels, we consider the following convolutional layers: 
\begin{itemize}
    \item $32$ channels and $D=W=112$.
    \item $64$ channels and $D=W=56$,
    \item $128$ channels and $D=W=28$, and
    \item $256$ channels and $D=W=14$
\end{itemize}

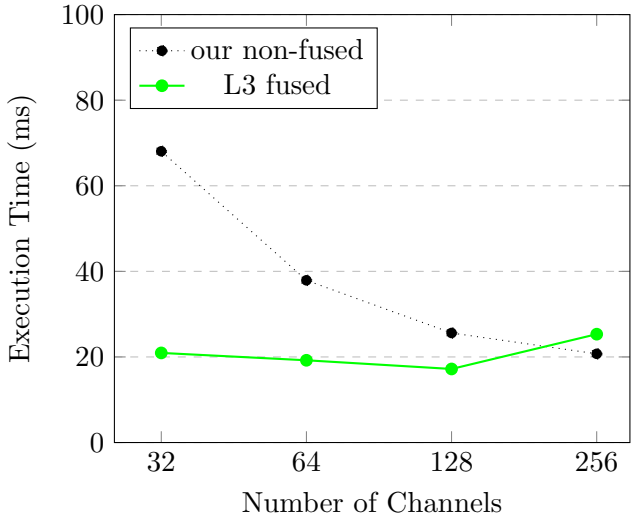
\begin{figure}
\begin{tikzpicture}
\begin{axis}[
    xmode=log,
    log ticks with fixed point,
    scaled x ticks=real:2,
    xlabel={Number of Channels},
    ylabel={Execution Time (ms)},
    xmin=0, xmax=300,
    ymin=0, ymax=100,
    xtick={32, 64,128,256},
    ytick={0,20,40,60,80,100},
    legend pos=north west,
    ymajorgrids=true,
    grid style=dashed,
]

 \addplot[
    color=black,
    mark=*,
    dotted,
    ]
    coordinates {
    (32, 68.06)(64, 37.91)(128, 25.6)(256, 20.73)
    };
    ]

 \addplot[
    color=green,
    mark=*,
    thick,
    ]
    coordinates {
    (32, 20.94)(64, 19.22)(128, 17.19)(256, 25.32)
    };
    ]

\legend{our non-fused, L3 fused}

\end{axis}
\end{tikzpicture}
\caption{Benchmark results on 4-core MacBook Pro}
\label{fig:mac}
\end{figure}

The results are presented in~\figureref{fig:mac} and validate our performance analysis for the i9 architecture. It shows that L3 fusion approach is widely applicable and provides impressive performance for interesting parameter regimes as long as a shared L3 cache is available.
\section{Conclusions}
\label{sec:conclusions}

While, there's no `one fits them all` approach, we advance the state-of-the-art by providing the most efficient algorithm for modern CPUs for executing many typical convolutional layers using transformed computation. Instead of computing full intermediate results, that are large and get stored in the main memory, our algorithm crucially relies on the shared L3 cache, which on modern processors is quite large, and quite likely could increase further in the future (e.g. see~\cite{crystalwell}). We show that the L3 cache pressure depends on the number of channels, thus, as the number of channels increases the L3-fused approach becomes inferior to the standard non-fused one. However, the trade-off is expected to swing more towards our fused approach for upcoming CPUs with larger shared caches.

In our work, we explained how to find a theoretically optimal value for the hyper-parameter $R$. This parameter can be tuned. Tuning $R$ is not hard, and can be done once and stored in a wisdom file.

It is an interesting future direction to apply the same ideas for other problems or for other types of hardware. Another question is to see whether in some cases it could be faster to compute a convolution with $C' = c_1 \cdot C$ input and $C'' = c_2 \cdot C'$ output channels by performing $c_1 c_2$ convolutions each with $C$ input and $C'$ output channels (and appropriately summing up the results), especially if L3-fusion can be super efficient for each of these smaller convolutions. Of course the trade-off here is the cost of performing input transforms $c_1$ times and output transforms $c_2$ times.

\bibliographystyle{abbrv}
\bibliography{biblio}
\end{document}